\documentclass[fleqn,usenatbib]{mnras}

\usepackage{newtxtext,newtxmath}
\usepackage[T1]{fontenc}

\DeclareRobustCommand{\VAN}[3]{#2}
\let\VANthebibliography\thebibliography
\def\thebibliography{\DeclareRobustCommand{\VAN}[3]{##3}\VANthebibliography}

\usepackage{graphicx}	
\usepackage{amsmath}	

\title[A model of the GSMF and UV LF at early times]{No evidence (yet) for increased star-formation efficiency at early times}

\author[C.\,T. Donnan et al.]{C.\,T. Donnan$^{1}$\thanks{E-mail: callum.donnan@ed.ac.uk},
J.\,S. Dunlop$^{1}$,
R.\,J. McLure$^{1}$,
D.\,J. McLeod$^{1}$,
F. Cullen$^{1}$
\\
$^{1}$Institute for Astronomy, University of Edinburgh, Royal Observatory, Edinburgh, EH9 3HJ, UK
}

\date{Accepted XXX. Received YYY; in original form ZZZ}

\pubyear{\the\year{}}

\begin{document}
\label{firstpage}
\pagerange{\pageref{firstpage}--\pageref{lastpage}}
\maketitle

\begin{abstract}

Early {\it JWST} observations have revealed substantial numbers of galaxies out to redshifts as high as $z \simeq 14$, reflecting a slow evolution of the galaxy UV luminosity function (LF) not anticipated by many models of galaxy evolution. The discovery of fairly massive galaxies at early times has again been viewed as a challenge to our understanding of early galaxy growth or even ${\rm \Lambda}$CDM cosmology. Here we develop and test a simple theoretical model which shows that these observations are unsurprising, but instead are arguably as expected if one assumes a non-evolving halo-mass dependent galaxy-formation efficiency consistent with that observed today. Crucially, this model matches the observed galaxy UV LF at $z \simeq 6-13$ {\it and} the galaxy stellar mass function (GSMF) at $z \simeq 6-8$. Using new constraints on Lyman continuum escape and the ionizing photon production efficiency, we also predict the progress of cosmic hydrogen reionization consistent with current observations. The requirement to fit both the UV LF and the GSMF breaks the degeneracy between mass-to-light ratio and star-formation efficiency, where the typical mass-to-light ratio of galaxies increases systematically with redshift beyond $z \simeq 6$. However, at present this does not require changes to the IMF, cosmic dust, or any other new astrophysics. Rather, the current data can be reproduced simply by assuming ever-younger stellar populations consistent with a formation epoch at $z \simeq 15$. A key prediction of our model therefore is that there should be a more rapid drop-off in the galaxy number density beyond $z \simeq 15$, where one can no longer appeal to ever younger ages to offset the precipitous descent of the halo mass function.
\end{abstract}


\section{Introduction}

In only its first two years of operation, {\it JWST} has already significantly advanced our knowledge of galaxy evolution back into the first 500 million years of cosmic time, producing a wealth of new observational results which have revolutionized, and in some cases challenged our understanding of early galaxy formation and growth.

Thanks to its sensitivity and extended near-infrared wavelength coverage (out to $\lambda\simeq5\, \mu \rm m$) the NIRCam instrument on \textit{JWST} has produced the first robust detections of significant numbers of galaxies at redshifts $z>10$ \citep[e.g.][]{castellano2022,naidu2022,donnan2023a, donnan2023b,adams2022, finkelstein2022c,harikane2023a}. This has in turn enabled the first robust measurements of the evolving galaxy UV luminosity function (LF) at $z=9-15$, revealing a perhaps surprisingly gradual evolution of the LF over this redshift range \citep[e.g.][]{finkelstein2023, perezgonzalez2023, mcleod2023, adams2023, donnan2024}. A number of the new high-redshift  galaxy candidates have also been spectroscopically confirmed using the NIRSpec instrument (now up to $z\simeq14.2$) \citep[e.g.][]{curtislake2022, bunker2023, arrabalharo2023, carniani2024} as well as ALMA \citep{schouws2024}. Although the relatively slow evolution of the galaxy population out to extreme redshifts is consistent with some pre-\textit{JWST} predictions from simulations \citep{wilkins2022} and observational extrapolations  \citep{mcleod2016}, it is undeniably in tension with a number of pre-\textit{JWST} models, particularly at the bright end \citep[e.g.][]{mason2015, tacchella2018, yung2019}.

Several alterations to the physical properties of galaxies at early times have been proposed in order to explain the high-redshift evolution of the galaxy UV LF. Firstly, changes in the star-formation efficiency as a function of redshift have been invoked. For example, \citet{harikane2023a} suggest that a constant efficiency model (based on that from \citet{harikane2022}) is unable to match the cosmic star-formation rate density ($\rho_{\rm SFR}$) at $z>10$. The implication is that an increase in star-formation efficiency with increasing redshift is required in order to match the observed prevalence of UV-luminous galaxies at $z\geq10$.

Secondly, much reduced dust attenuation has been suggested to explain the bright-end of the UV LF at $z=11$ \citep[e.g.][]{ferrara2023} with it being proposed that any dust created in early galaxies could possibly (at least temporarily) be ejected due to the extreme physical conditions associated with early galaxy formation. A lack of dust at extreme redshifts is also consistent with measurements of the UV continuum slopes of galaxies at $z=11$ (derived from \textit{JWST}/NIRCam photometry) which are, on average, consistent with $\beta\simeq-2.5$  \citep[e.g.][]{topping2023,morales2023,cullen2023}.

Another fundamental scaling relation is the galaxy stellar mass function (GSMF) which directly traces the product of star formation across cosmic time. By measuring the GSMF, one can obtain a direct probe of past star formation which, given the relatively short timescales involved at high redshifts, can be related to star-formation efficiency as a function of dark matter halo mass. Precise measurement of the GSMF requires photometry of the rest-frame optical light from galaxies in order to provide accurate stellar mass measurements for individual galaxies. Using \textit{HST} as well as ground-based observations, the GSMF was previously measured up to $z\simeq 3$ by a number of studies \citep[e.g.][]{muzzin2013,mortlock2015, mcleod2021,weaver2023}. However, at $z\geq3$, measurements of the GSMF proved more challenging because, prior to the launch of \textit{JWST}, this could only be achieved using relatively low resolution \textit{Spitzer}/IRAC observations to probe the rest-frame optical at higher redshifts \citep[e.g.][]{grazian2015,song2016,stefanon2021}. However, now \textit{JWST} has enabled a number of new measurements of the GSMF, with the sensitivity and exquisite angular resolution of NIRCam at $\lambda\simeq1-5\mu$m enabling the first robust measurements of the GSMF over the redshift range $z=3-8$ \citep{weibel2024,harvey2024,shuntov2024}. The discovery of at least some very massive galaxies at these early times has again led to the suggestion that increases in star-formation efficiency are required to explain their existence \citep[e.g. $\epsilon\sim0.2$;][]{chworowsky2024}, with some authors even suggesting that the most extreme examples present a challenge to the standard $\rm \Lambda$CDM cosmological model \citep[e.g. $\epsilon\sim0.8-1.0$;][]{labbe2023,glazebrook2024,shuntov2024}. 

Finally, there have been suggestions that there is an ionizing photon budget crisis \citep{munoz2024} in the sense that some recent \textit{JWST} results appear to imply that too many ionizing photons were produced at early times to be consistent with measurements of the evolving neutral fraction of the intergalactic medium \citep[e.g.][]{tang2024,nakane2024} including the requirement that reionization is not fully completed until $z\simeq5.5$ \citep{bosman2021}.

Given the recent advances in the observational determinations of both the UV LF and the GSMF, it is clearly important to generate a theoretical framework that can simultaneously explain the form and evolution of both of these fundamental scaling relations. In this paper, we expand upon and further develop the model introduced in \citet{donnan2024} to explore whether a simple theoretical model can reproduce the \textit{JWST} constraints on the UV LF, the GSMF, and the timeline of reionization at $z=6-13$. This can be viewed as a test of the extent to which current observational data really do require the introduction of more complex/changing astrophysics to explain the new results on early galaxy evolution emerging from \textit{JWST}.

The paper is structured as follows. In Section \ref{sec:GSMF} we initially model the GSMF and compare to the latest observational constraints. In Section \ref{sec:LF} we further develop the model to generate the UV LF at $z=6-13$. In Section \ref{sec:dust_attenuation} we then introduce dust attenuation to our model of the UV LF and present the implied UV attenuation$-$stellar mass relation. In Section \ref{sec:reionization} we then explore the predictions of our model for the progress of cosmic hydrogen reionization. In Section \ref{sec:discussion} we then discuss the implications of our theoretical model for star-formation efficiency, high-mass feedback, dust attenuation and the ionizing photon budget. Finally, in Section \ref{sec:conclusions} we summarize our conclusions. Throughout we use magnitudes in the AB system \citep{oke1974,oke1983}, a \citet{chabrier2003} initial mass function, and assume a standard cosmological model with $H_0=70$ km s$^{-1}$ Mpc$^{-1}$, $\Omega_m=0.3$ and $\Omega_{\Lambda}=0.7$.

\section{Modelling the stellar mass function}
\label{sec:GSMF}
\subsection{The star-formation efficiency}
We begin by constructing a model of the galaxy stellar mass function (GSMF) which we describe in this section. 
To construct this simple model we first calculate the evolving dark-matter halo mass function (HMF) using \textsc{HMFCalc} \citep{murray2013} where we use the model of \citet{reed2007}. A number of alternative models of the HMF are available which scatter across a range of $\sim0.15$ dex at $z=0$. This scatter increases with redshift with a $\sim0.35$ dex scatter at $z=10$ rising to a scatter of  $\sim0.45$ dex at $z=13$. We adopt the \citet{reed2007} model specifically because this is lies in the centre of this scatter over the halo mass range of concern. The \citet{reed2007} model was also specifically determined by simulating halos at $z=10-30$ making it ideal for this work. 

\begin{figure}
	\includegraphics[width=\columnwidth]{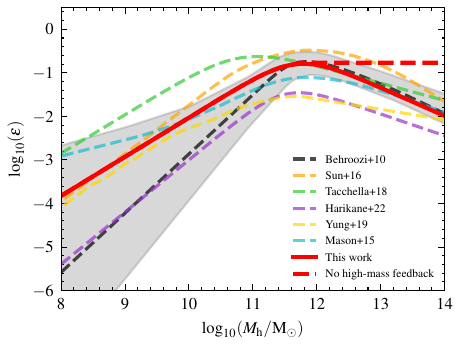}
    \caption{The adopted (redshift independent) star-formation efficiency ($\epsilon$) as a function of halo mass in our model from \citet{donnan2024} (solid red line) compared to the corresponding relations in the models from \citet{sun2016}, \citet{tacchella2018}, \citet{harikane2022}, \citet{yung2019} and \citet{mason2015}. The dashed grey line shows the $z~=~0.1$ \citet{behroozi2010} stellar-to-halo mass relation (SHMR) which is identical to the relation adopted here at high masses. The low-mass form of our adopted relation is chosen to better track the observational constraints on the SHMR at $z=0$ \citep{wechsler2018}, the uncertainty in which is shown here by the gray shaded region. The red dashed line shows our alternative adopted star-formation efficiency ($\epsilon$) model with no high-mass feedback, a model which we explain further in Section~\ref{sec:remove_feedback} but ultimately reject in Section~\ref{sec:LF}.}  
    \label{fig:SHMR}
\end{figure}

We model the star-formation efficiency as a function of halo mass, $\epsilon(M_{\rm h})$ through the stellar-to-halo mass relation (SHMR). This relates the stellar and halo masses, incorporating the universal baryon fraction, $f_{\rm b}=0.167$, through the relation

\begin{equation}
    \frac{M_{*}}{M_{h}} = \epsilon(M_{\rm h}) f_{\rm b}
\end{equation}

\noindent
which assumes that the efficiency, $\epsilon$, is purely dependent on the mass of a galaxy's host dark-matter halo. We then adopt a functional form for $\epsilon(M_{\rm h})$ using the double power-law relationship from \citet{tacchella2018}:

\begin{equation}
\label{eq:dpl_eff}
    \epsilon(M_{\rm h}) = 2\epsilon_0 \left[ \left(\frac{M_{\rm h}}{M_{\rm c}}\right)^{-\beta} + \left(\frac{M_{\rm h}}{M_{\rm c}}\right)^{\gamma} \right]^{-1}
\end{equation}

\noindent
where $\epsilon_0$ is the peak efficiency, $M_{\rm c}$ is the characteristic mass (the mass at $\epsilon_0$), $\beta$ is the low-mass slope and $\gamma$ is the high-mass slope. We initially use $(\epsilon_0$, $M_{\rm c} / \rm M_{\odot}$, $\beta$, $\gamma) = (0.16, 10^{11.7}, 0.9, 0.65)$, as this is the parameterisation used in \citet{donnan2024}. This relation is shown by the solid red line in Fig.~\ref{fig:SHMR}. We also show in Fig.~\ref{fig:SHMR} the observational constraints on the $z \simeq 0$ SHMR as compiled by \citet{wechsler2018} with the uncertainty captured by the gray shaded region.

We then also compare our initial adopted relation to the predictions from a number of alternative models. These alternative models cover a large range of efficiencies and include some that are clearly inconsistent with the $z=0$ relation. It can be seen that \citet{tacchella2018} assume a significantly greater efficiency ($\sim 1\, \rm{dex}$) whereas the model from \citet{mason2015} is moderately lower and the models from \citet{sun2016} and \citet{harikane2022} are significantly lower. For the semi-analytical model from \citet{yung2019}, we plot their $z=10$ relation. Another feature of these models is that they assume a rather low efficiency for massive dark-matter halos ($M_{h}>10^{11.7}\, \rm M_{\odot}$) compared to both our initially adopted model and indeed the relation at $z=0$.  These models had been specifically designed to model the UV LF at $z=4-10$. However, the UV LF is not a direct tracer of star-formation efficiency as the UV luminosity of a galaxy is dictated by more than simply its stellar mass (e.g. stellar age, star-formation history). Therefore, although these models all assume significantly different star-formation efficiencies, they often still yield similar final UV LFs at $z=4-10$ due to these degeneracies. This is important to note since, as discussed later in Section~\ref{sec:cst_SF}, these models have led to claims that a constant star-formation efficiency model fails to reproduce the now well-constrained UV LF at $z>10$ \citep[e.g.][]{harikane2023a,tacchella2024}. However, we re-emphasize here that, as demonstrated in Fig.~\ref{fig:SHMR}, there are in fact very substantial differences in what form any proposed constant star-formation efficiency is assumed to take.

\subsection{Observational constraints on the GSMF}
Given that there are such significant differences between the star-formation efficiency relations assumed by different models, breaking the degeneracy between any evolution in the efficiency function and, for example, changing mass-to-light ratios in galaxies requires direct tests of the evolving GSMF predicted by the different models. Using the $\epsilon(M_{\rm h})$, halo mass can be converted to a stellar mass, $M_{\rm *}$, by:
\begin{equation}
    M_{\rm *} = \epsilon(M_{\rm h}) f_{\rm b} M_{\rm h}
\end{equation}

\noindent
The galaxy number density as a function of stellar mass is then given by:
\begin{equation}
    \frac{dn}{d \log(M_{\star})} = \frac{d \log(M_{\rm h})}{d \log(M_{\star})} \times \frac{dn}{d \log(M_{\rm h})}
\end{equation}

\noindent
Therefore, by assuming a form for the HMF, we can derive a model of the GSMF using a functional form for $\epsilon(M_{h})$. This allows measurements of the GSMF to directly establish the star-formation efficiency, hence enabling direct tests of our model as well as those presented in the literature.

\begin{figure*}
	\includegraphics[width=\textwidth]{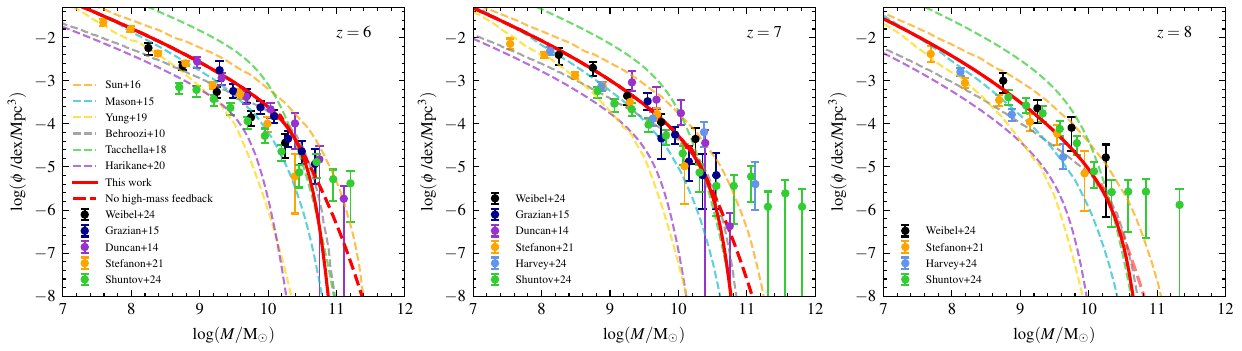}
    \caption{Our model predictions for the evolving galaxy stellar mass function (GSMF) at $z=6-8$ based on the two alternative models of $\epsilon(M_h)$ presented in Fig.~\ref{fig:SHMR}) are shown in each panel by the red lines (where the dashed version  indicates the removal of high-mass feedback). Also shown by data-points identified in the legend are the latest observational measurements of the GSMF from \citet{duncan2014,grazian2015,stefanon2021,weibel2024,harvey2024,shuntov2024}. The predictions of the GSMF using the alternative $\epsilon(M_h)$ relations shown in  Fig.~\ref{fig:SHMR} from \citet{tacchella2018}, \citet{harikane2022}, \citet{yung2019}, \citet{sun2016}, \citet{mason2015} and \citet{behroozi2010} are shown by the various fainter coloured dashed lines for comparison. All stellar masses have been adjusted to assume a \citet{chabrier2003} initial mass function.}  
    \label{fig:SMF}
\end{figure*}

With sensitive imaging at $\lambda>2\mu$m, \textit{JWST} now allows robust measurement of the GSMF at $z=6-8$. In Fig.~\ref{fig:SMF} we compare our model prediction for the evolving GSMF at $z=6-8$ to the latest observational measurements from a number of pre-\textit{JWST} \citep{duncan2014,grazian2015,stefanon2021} and now \textit{JWST} measurements \citep{harvey2024,weibel2024,shuntov2024}.
This demonstrates that our model is able to closely match the observational data points of the GSMF at $z=6-8$. This also confirms that the simple assumption of a star-formation efficiency function unchanged from that observed at $z=0$ is sufficient to successfully reproduce the GSMF out to the highest redshifts for which robust measurements have so far been achieved. Crucially, however, Fig.~\ref{fig:SMF} also demonstrates that several of the other models in the published literature, while designed to potentially reproduce the $z=4-8$ UV LF, clearly fail this fundamental GSMF test.

\subsection{The effect of removing high-mass feedback}
\label{sec:remove_feedback}
Although the star-formation efficiency function used in \citet{donnan2024} is successfully able to broadly match the observational measurements of the GSMF at $z=6-8$, this $\epsilon(M_{\rm h})$ was designed primarily to reproduce the UV LF at $z\geq9$ where the halo masses of interest lie below the peak of the star-formation efficiency relation (i.e. $\log(M_{\rm h}/ \mathrm{M_{\odot}})\lesssim11.7$). Therefore, in that work we simply assumed a high-mass slope equal to that of the $z=0.1$ SHMR from \citet{behroozi2010} of $\gamma = 0.65$. However, at $z\geq7$ there have been suggestions that it may make sense to invoke a lack of high-mass feedback at early times to explain the excess in the bright-end of the UV LF (compared to a Schechter function) through increased star-formation rates at a given halo mass \citep[e.g.][]{bowler2017,bowler2020,stefanon2019}. Given that the GSMF is a more direct probe of the star-formation efficiency and now that it has been better determined at $z=6-8$, we are able to test if the high-redshift form of the GSMF does indeed require 
less high-mass feedback than at later times. To explore this we introduce an alternative form of $\epsilon(M_{\rm h})$ in which a constant value of $\epsilon(M_{\rm h}) = 0.17$ is assumed for $\log(M_{\rm h}/ \mathrm{M_{\odot}})>11.7$, thus essentially removing all feedback within high-mass halos. This modification is illustrated in Fig.~\ref{fig:SHMR} by the dashed red line  which, for simplicity, has the same functional form for $\epsilon(M_{\rm h})$ as our original version at lower masses, $\log(M_{\rm h}/ \mathrm{M_{\odot}})<11.7$.

From Fig.~\ref{fig:SMF} it can be seen that, for the majority of the stellar mass-range probed by the observational measurements of the GSMF, the removal of high-mass feedback has no effect because these observations probe modest halo masses of $\log(M_{\rm h}/ \mathrm{M_{\odot}})<11.7$. However, the highest-mass bins at $z=7$ in particular are more closely matched by our model with high-mass feedback removed. Therefore, one might conclude that there is tentative evidence in favour of the removal of high-mass feedback at $z=7$. 

However, caution is required when comparing to observations of the high-mass end of the GSMF. Firstly, the high-mass end of the GSMF is vulnerable to Eddington bias \citep{eddington1913}. When measuring the GSMF, each galaxy in the sample has a stellar mass measurement with an associated uncertainty. This means that when measured, galaxies scatter above and below the true stellar mass value. Since there are a lot more low-mass galaxies than high-mass galaxies (as the GSMF declines steeply at high mass) then more of the low-mass galaxies scatter upwards than  high-mass galaxies scatter downwards, resulting in an artificial boosting of the number densities of galaxies in the high-mass bins.

Secondly, the galaxy number densities in the highest-mass bins are inevitably highly uncertain because they are often based on a very small number of galaxies, or in extreme cases even the proposed presence of a single high-mass source. This is particularly true for the high-mass bins from \citet{shuntov2024}. Therefore, even a small number of contaminants (e.g. due to erroneous photometric redshifts) can drastically change the measured number density at $\log(M_{\star}/ \mathrm{M_{\odot}})\gtrsim10.6$. 
Ultimately, there are not yet enough confirmed galaxies with $\log(M_{\star}/ \mathrm{M_{\odot}})\gtrsim10.6$ at $z\geq6$ to place meaningful constraints on the GSMF in this mass range. Increased survey areas with enhanced filter coverage are required to robustly test the need for high-mass feedback through direct comparison with the high-redshift form of the GSMF. In Section~\ref{sec:LF} we present another test of the need for high-mass feedback using our model of the intrinsic UV luminosity function. 

\section{The galaxy UV luminosity function}
\label{sec:LF}
\subsection{Observational measurements of the UV LF at $\mathbf{z=6-13}$}
The next stage of the model is to convert the GSMF into a galaxy UV luminosity function (LF) in order to compare our model to observational measurements of the UV LF at $z=6-13$. 

The UV LF at $z\geq6$ has primarily been measured using photometric surveys from space-based telescopes (e.g. \textit{HST}, \textit{JWST}). Prior to the launch of \textit{JWST}, there existed a general consensus over the measured form of the UV LF at $z=6-8$, as described by an evolving Schechter function with a fairly steep faint-end slope of $\alpha \simeq -1.8$ \citep[e.g.][]{bouwens2007,mclure2009,mclure2010,bouwens2021}. However, \textit{HST} data alone were unable to probe the bright-end of the UV LF ($M_{\rm UV}\lesssim-21.5$) due to the limited area of \textit{HST} surveys. Fortunately, wider-area ground-based surveys were able to probe this luminosity range, hence enabling measurements of the bright-end of the UV LF, at least out to $z \simeq 9$. 

Therefore, in this work we include the latest ground-based constraints on the bright-end of the UV LF in combination with a representative fainter dataset from space-based surveys at each redshift. This forms the dataset that we use to model the UV LF. At $z=6$ we use the \textit{HST} measurements from \citet{bouwens2021} with the ground-based data from \citet{bowler2015}. At $z=7$ we use the data from \citet{mclure2013} in combination with that provided by \citet{bowler2017} and \citet{varadaraj2023}. At $z=9$ we use the data from \citet{mclure2013} combined with that from \citet{bowler2020} and \citet{donnan2023a}. At $z\geq9$ we use the \textit{JWST} measurements from \citet{donnan2024} including bright-end constraints at $z=9$ from \citet{bowler2020} and \citet{donnan2023a}. 

\subsection{Modelling the UV LF}
To convert our model of the GSMF into a predicted UV LF, we assume a redshift-dependent mass-to-light ratio given by a linear relationship between $M_{\rm UV}$ and $M_{\star}$:
\begin{equation}
    M_{\rm UV} = M_{\rm UV, *} -2.5 \left[\log_{10}(M_{\star} / \mathrm{M_{\odot}}) - 9 \right]
\end{equation}

\noindent
where $M_{\rm UV, *}$ is the rest-frame UV absolute magnitude corresponding to $\log_{10}(M_{\star} / \mathrm{M_{\odot}})=9$. This relationship is therefore defined by the UV luminosity which is assumed to be produced by a star-forming galaxy with a stellar mass of $M_{\star} = 10^9  \mathrm{M_{\odot}}$. To determine this we perform a $\chi^2$ minimization fit to the faint-end of the UV LF data points ($M_{\rm UV}>-19$) where, at $z=6-13$, the galaxy population can be reasonably expected to be virtually dust-free and therefore the observed UV flux density from these fainter galaxies can be assumed to be un-attenuated by dust. The resultant $M_{\rm UV, *}$ required for our model to reproduce the UV LF as a function of redshift is noted in the second column of Table \ref{tab:ages_model}. The $M_{\rm UV}$ for a galaxy of a given stellar mass can then be linked to the age of the stellar population within that galaxy, since a younger stellar population at fixed stellar mass has an increased UV luminosity and hence a more negative absolute magnitude. 

\begin{figure}
	\includegraphics[width=\columnwidth]{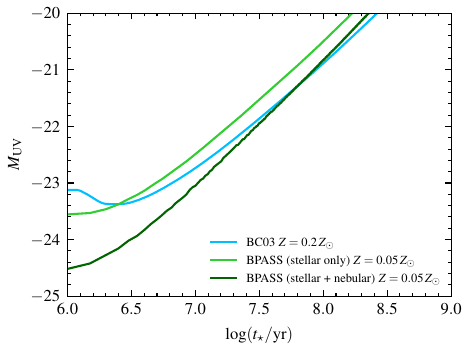}
    \caption{Predicted galaxy UV absolute magnitude as a function of stellar mass-weighted age for a BC03 stellar population model \citep{bruzual2003} with a constant star-formation history and a  metallicity of $Z=0.2 Z_{\odot}$ (blue), alternatively for a BPASS model \citep[][]{Eldridge2017} with a constant star-formation history and a metallicity of $Z=0.05 Z_{\odot}$ with a stellar continuum (light green) and stellar + nebular continuum (dark green).}  
    \label{fig:sp_comparison}
\end{figure}

To determine the stellar age associated with the required $M_{\rm UV}$ for a stellar mass of $\log_{10}(M_{\star}/ \mathrm{M_{\odot}})=9$ at each redshift, we employ stellar population synthesis codes. In this approach we assume that at a given redshift the average age of a galaxy is independent of it's stellar mass. We use two different codes in this work in order to explore the feasibility of and uncertainty in our approach. First we use a BC03 stellar population model \citep{bruzual2003} with a constant star-formation history and a metallicity of $Z=0.2 Z_{\odot}$. Second, we use The Binary Population and Spectral Synthesis code \citep[BPASS;][]{Eldridge2017} with a constant star-formation history but with a lower metallicity of $Z=0.05 Z_{\odot}$. We determine this initially with the stellar continuum only before re-calculating the model with an added contribution of the nebular continuum. The relationships between UV magnitude and the stellar mass-weighted age for a galaxy of $\log_{10}(M_{\star} / \mathrm{M_{\odot}})=9$ for each model are shown in Fig.~\ref{fig:sp_comparison}. This demonstrates that, for galaxies with ages $\gtrsim 3 \, \rm Myr$, the stellar only BPASS model consistently produces fainter UV magnitudes whereas this comparison reverses for ages $\lesssim 5 \, \rm Myr$ for the stellar only model. Alternatively, for a given UV magnitude, the BPASS model produces younger stellar ages for $M_{\rm UV}\gtrsim -23$. For the BPASS model with the addition of nebular continuum, the ages are similar to the BC03 model at $M_{\rm UV}\gtrsim -23$ but allow moderately older ages for $M_{\rm UV}\lesssim -23$. For the $M_{\rm UV, *}$ at each redshift, the BC03 and BPASS ages are noted in the latter two columns in Table~\ref{tab:ages_model}. 

\begin{table}
	\centering
	\caption{The age-dependent UV magnitude mapping to stellar mass as a function of redshift for our model. The first column is the redshift. The second column is the UV absolute magnitude, $M_{\mathrm{UV}}$, that corresponds to a stellar mass of $\log(M_*/{\rm M_{\odot}})=9$ determined by the mass-weighted age given in the third column for a BC03 stellar population model \citep{bruzual2003} with a metallicity of $Z/{\rm Z_{\odot}=0.2}$. The fourth column is the inferred mass-weighted age for a BPASS \citep[][]{Eldridge2017} model with metallicity of $Z/{\rm Z_{\odot}=0.05}$ with only a stellar continuum. Finally, the fifth column is the mass-weighted age for a BPASS \citep[][]{Eldridge2017} model with metallicity of $Z/{\rm Z_{\odot}=0.05}$ with both a stellar and nebular contribution to the continuum.}
	\label{tab:ages_model}
    \setlength{\tabcolsep}{4pt} 
	\renewcommand{\arraystretch}{1.15} 
	\begin{tabular}{lcccc} 
		\hline
         & & Age /Myr & Age /Myr & Age /Myr\\
		z & $M_{\mathrm{UV}}$ /AB mag & BC03 & BPASS St. only & BPASS St. + Neb.\\
		\hline
  6 & $-20.59$ & 140 & 91 & 128 \\
  7 & $-20.99$ & 89 & 58 & 85 \\
  8 & $-21.03$ & 86 & 56 & 82 \\
  9 & $-21.14$ & 75 & 49 & 74 \\
  10 & $-21.76$ & 36 & 24 & 39 \\
  11 & $-22.74$ & 10 & 7 & 14 \\
  12.5 & $-23.27$ & 2.5 & 3 & 8 \\
  \hline
	\end{tabular}
\end{table}

\subsection{Predictions for high-redshift sub-millimeter galaxies}
\label{sec:submm}
\begin{figure*}
	\includegraphics[width=\textwidth]{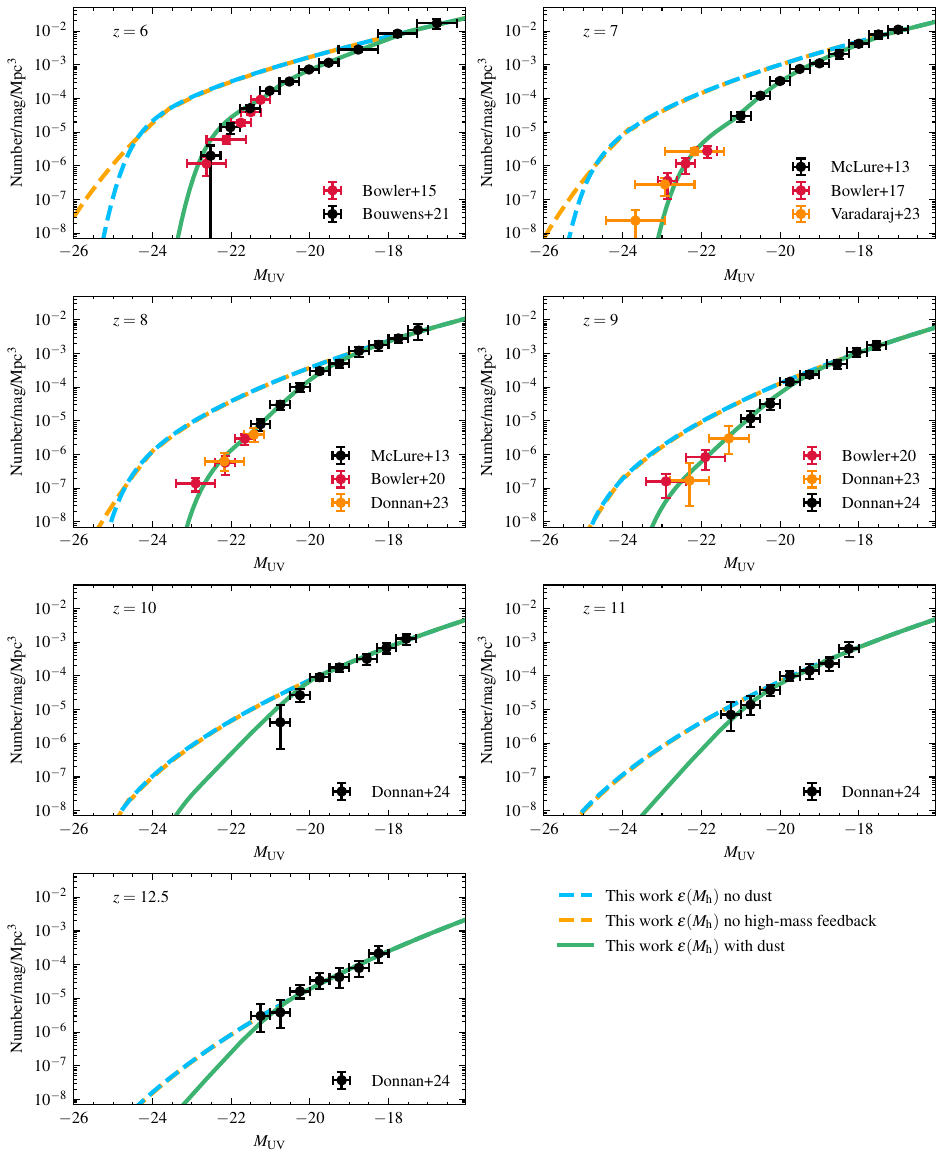}
    \caption{A comparison of our models of the UV luminosity function (UV LF) to observational measurements from \citet{mclure2013,bowler2015,bouwens2021,varadaraj2023,donnan2023a,donnan2024}. The dashed blue line shows our intrinsic dust free model assuming the form of star-forming efficiency $\epsilon (M_{\rm h})$ established by \citet{donnan2024}. The dashed orange line shows the dust-free model assuming the \citet{donnan2024} $\epsilon (M_{\rm h})$ model is altered to contain no high-mass feedback. The solid green line shows the final model with dust attenuation.}   
    \label{fig:UVLF}
\end{figure*}

Using our model of the intrinsic UV LF (i.e. without any dust attenuation, as shown in Fig.~\ref{fig:UVLF} at $z = 6, 7, 8, 9, 10, 11, 12.5$)
we perform another test of the evidence for and against high-mass feedback in high-redshift galaxies. In this case we estimate the number of heavily dust-obscured galaxies that would be detectable by sub-mm surveys under either scenario. These are galaxies that contain significant levels of dust attenuation making them extremely faint in the UV and therefore potentially absent from observations of the bright end of the UV LF. However, any such highly star-forming galaxies which are largely attenuated in the rest-frame UV should be revealed at sub-mm wavelengths due to the re-emission of the absorbed UV light by warmed dust grains in the rest-frame far-infrared.

To estimate the prevalence of bright sub-mm sources at high redshifts, inferred from the need to effectively remove them from the bright end of the UV LF through dust obscuration, we proceed as follows. 

For simplicity, we consider only galaxies at $z=6-8$ which our model predicts to have an intrinsic unobscured UV absolute magnitude brighter than $M_{\rm UV}=-24.98$. This choice was made in order to capture the number density of galaxies which have a SFR $\ge 300\, \rm M_{\odot} yr^{-1}$ according to the conversion factor given by \citet{madau2014}. Specifically, we use the relation:

\begin{equation}
    \mathrm{SFR} = \cal{K}_{\rm UV} L_{\rm UV}
\end{equation}

\noindent
where $\cal{K}$$_{\rm UV} = 0.70 \times 10^{-28}$ M$_{\odot}$ yr$^{-1}$/erg s$^{-1}$ Hz$^{-1}$ (adjusted to a \citet{chabrier2003} initial mass function) and $L_{\rm UV}$ is the UV luminosity. 

Next, we exploit the work of \citet{dunlop2017} who derived a relationship between the infrared-visible SFR and observed flux density at $\lambda = 1.3 \rm \, mm$ ($S_{1.3}$) by fitting the characteristic spectral energy distribution (SED) of the ALMA-detected sources in the HUDF across the wavelength range $24 \,\mathrm{\mu m}-1.3 \,\mathrm{mm}$ giving:
\begin{equation}
    \mathrm{SFR / M_{\odot}yr^{-1}} \simeq 0.3 S_{1.3}/ \mu \rm Jy 
\end{equation}

We then simply assume that the intrinsic UV-derived $\rm SFR$ is equal to the actual infrared-inferred $\rm SFR$ to determine the flux density at $1.3\, \rm mm$ (a very reasonable assumption for highly-obscured star-forming galaxies). For $\mathrm{SFR}\simeq300\, \rm M_{\odot} yr^{-1}$ we derive a $1.3\, \rm mm$ flux-density estimate of $S_{1.3}\simeq 1.0\, \rm mJy$. We then assume a functional form for the shape of the far-IR SED such that $f_{\nu} \propto \nu^3$ to determine the equivalent flux density at $\lambda_{\rm obs}= 850\, \rm \mu m$. This results in a value of $S_{850}\simeq 3.6\, \rm mJy$ which enables comparison to SCUBA-2 surveys which have typically achieved 4-$\sigma$ detections and hence published source catalogues down to this $850\, \rm \mu m$ flux density \citep[e.g. S2COSMOS, AS2UDS;][]{simpson2019,dudzeviciute2020}.

Finally, we then integrate our model of the intrinsic UV LF at $z=6-8$ to a limit of $M_{\rm UV}=-24.98$ to determine the number of galaxies that would be detectable in a $850\, \rm \mu m$ survey down to a flux density of $S_{850}=3.6\, \rm mJy$. 

We compare our model predictions to the AS2UDS survey which detected 364 SMGs across $\sim1 \, \rm sq.\, degrees$ in the UDS field with $S_{\rm 850\mu m} \ge 3.6 \rm mJy$ based on ALMA follow-up of the original SCUBA-2 Cosmology Legacy Survey of the UDS field \citep{dudzeviciute2020}. They report that 2 of these sources at $z>5.5$ indicating a number density of $\sim2$ galaxies per sq. degree with $S_{\rm 850\mu m}=3.6 \rm mJy$ at $z>5.5$. This is consistent with the more recent \textit{JWST} analysis of the S2COSMOS sample in COSMOS-Web \citep{caset2023b} which revealed 2 sources (at $z=5.5-8.5$ ) across $0.54 \, \rm sq.\, degrees$ with a flux density $S_{850}>2\, m \rm Jy$ \citep{mckinney2024}. In this case these two sources have been specifically noted in the literature with AzTECC71 at $z_{\rm phot}=5.7 \pm 0.6$ \citep{mckinney2023} and MAMBO-9 at $z_{\rm spec}=5.850$ \citep{casey2019}. Their measurements therefore suggest $\sim3-4$ sources with $S_{850}>2\, \rm mJy$ per sq. degree. Given the somewhat increased depth used in this study compared to our calculation and the presence of an over-density at $z=6$ in which one of these sources apparently resides \citep{brinch2024}, it is consistent with the number density from the AS2UDS survey. We note however, that both the proposed $z > 5.5$ sources in the A2UDS survey only have photometric redshifts, and that our own photometric redshift analysis of this same sample in fact yields no sources with $z > 5.5$. Meanwhile, of the 2 claimed $z > 5.5$ sources in S2COSMOS sample, only the redshift of AzTECC71 can be considered secure. It is thus reasonable to adopt an estimate of 1 source per sq. degree with $z > 5.5$ at the $S_{\rm 850\mu m}=3.6 \rm mJy$ level.

Our model, including continued high-mass feedback (as seen at low redshift) at high redshift yields a prediction of $\simeq0.3$ galaxies with $S_{850}>3.6\, \rm mJy$ per sq. degree, completely consistent with the available sub-mm survey results. However, with high-mass feedback removed, we predict $\simeq4$ galaxies with $S_{850}>3.6\, \rm mJy$ per sq. degree, less consistent with the data. Accepting the unavoidable crudeness of this comparison, this demonstrates that the number density of bright sub-mm galaxies can be used to place constraints on the star-formation efficiency in high-mass halos ($M_{\rm h}\gtrsim10^{11.7}$) at early times. In this case the available evidence continues to support the necessity of high-mass feedback, at least out to $z \simeq 6$, or at the very least provides no motivation for increasing efficiency at high masses.
 
\section{Dust attenuation}
\label{sec:dust_attenuation}
Given that the intrinsic, dust-free, model is able to successfully reproduce submillimeter number counts, we now consider whether the introduction of a physically plausible relationship between stellar mass and dust attenuation can accurately reproduce the observed UV LF. In \citet{donnan2024}, we accounted for the impact of dust by assuming the $A_{1500}-M_{\star}$ relation derived for star-forming galaxies at $2<z<3$ by \citet{mclure2018}. In this study, we attempt to directly measure the $A_{1500}-M_{\star}$ relation as a function of redshift by forcing agreement between the intrinsic, dust-free, UV LF model and the observed data (as shown in Fig.~\ref{fig:UVLF}).

\begin{figure}
	\includegraphics[width=\columnwidth]{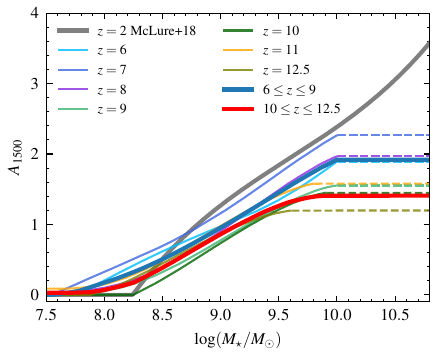}
    \caption{The UV dust attenuation at $\lambda_{\rm rest}=1500\,$\AA\, as a function of stellar mass derived in our model at each redshift. The $z=2$ relation from \citet{mclure2018} is shown by the solid gray line. The dashed lines show the horizontal extrapolation at high stellar masses. The thick solid blue line shows the mean $A_{1500}-M_{\star}$ relation at $6\leq z\leq 9$ whereas the thick solid red line shows the mean $A_{1500}-M_{\star}$ relation at $10\leq z\leq 12.5$. }  
    \label{fig:dust_mass}
\end{figure}

To determine the level of UV attenuation as a function of stellar mass, we calculate the difference between the intrinsic model of the UV LF and the observational data at the corresponding number density. The resulting predictions for the $A_{1500}-M_{\star}$ relation as a function of redshift are shown in Fig.~\ref{fig:dust_mass}, along with the $z \simeq 2.5$ relation from \citet{mclure2018}.

Due to the DPL functional form used to fit the observed UV LF, our adopted method produces a physically unrealistic turn-over in the derived $A_{1500}-M_{\star}$ relation at stellar masses of $\log(M_{\star}/ \rm M_{\odot})\simeq 10$, corresponding to absolute UV magnitudes well beyond the bright end of the observed UV LF data.
Consequently, at each redshift, we implement a horizontal extrapolation at stellar masses greater than the peak of the derived $A_{1500}-M_{\star}$ curve. 
This extrapolation, shown as the dashed lines in Fig.~\ref{fig:dust_mass}, begins at $\log(M_{\star}/ \rm M_{\odot})\simeq10$ for redshifts $z=6-9$ and at $\log(M_{\star}/ \rm M_{\odot})\simeq 9.9, 9.8, 9.6$ for $z=10, 11$ and 12.5, respectively. The final UV LF models, incorporating the derived dust attenuation at each redshift, are plotted as the solid green lines in Fig.~\ref{fig:UVLF}. By design, there is good agreement between the models and the observational data points across the range of UV magnitudes constrained by the available data. 

The individual redshift relations for $A_{1500}-M_{\star}$ shown in Fig.~\ref{fig:dust_mass} do not display a clear monotonic evolution with redshift although, at fixed stellar mass, there is a tentative trend towards lower dust attenuation at earlier times. This trend is made clearer when considering the mean $A_{1500}-M_{\star}$ relations in the range $6\leq z \leq9$ and $10\leq z \leq 12.5$, shown as the solid blue and red curves, respectively. Given the uncertainties involved, it is difficult to draw any firm conclusions on the functional form of the derived $A_{1500}-M_{\star}$ relations. However, they appear physically plausible and, at least within the stellar-mass range that is best constrained by the available data (i.e. $8.5 \leq \log(M_{\star}/ \mathrm{M_{\odot}}) \leq 9.5)$, have a similar gradient to the 
$z \simeq 2.5$ relation from \citet{mclure2018} which was derived using a completely different method.

\section{Cosmic hydrogen reionization}
\label{sec:reionization}
Constraining the timeline of reionization continues to be a key goal of extragalactic astronomy. Although there is general consensus that reionization was completed by $z\simeq5.5-6$ \citep{robertson2015,bosman2021}, the evolution of the neutral hydrogen fraction ($X_{\rm HI})$ at higher redshifts is still open to question.
Fundamentally, the timeline of reionization is governed by the balance between the rate at which ionizing photons are produced and escape into the IGM and the hydrogen recombination rate. The rate at which ionizing photons escape into the IGM is given by:
\begin{equation}
    \label{eq:nion}
    \Dot{n}_{\rm ion} = f_{\rm esc} \xi_{\rm ion} \rho_{\rm UV}
\end{equation}
\noindent
where $f_{\rm esc}$ is the escape fraction of Lyman continuum radiation, $\xi_{\rm ion}$ is the ionizing photon production efficiency and $\rho_{\rm UV}$ is the UV luminosity density. Consequently, by performing a luminosity-weighted integral of our UV LF models to measure $\rho_{\rm UV}$, it is possible to estimate $\Dot{n}_{\rm ion}$ as a function of redshift, based on assumed values of $f_{\rm esc}$ and $\xi_{\rm ion}$. When calculating $\rho_{\rm UV}$, we integrate our UV LF models to a common limit of $M_{\rm UV}=-13$, as is typically done when estimating the contribution of the growing galaxy population to cosmic hydrogen reionization \citep[e.g.][]{robertson2015,munoz2024,whitler2025}.

\begin{figure*}
	\includegraphics[width=\textwidth]{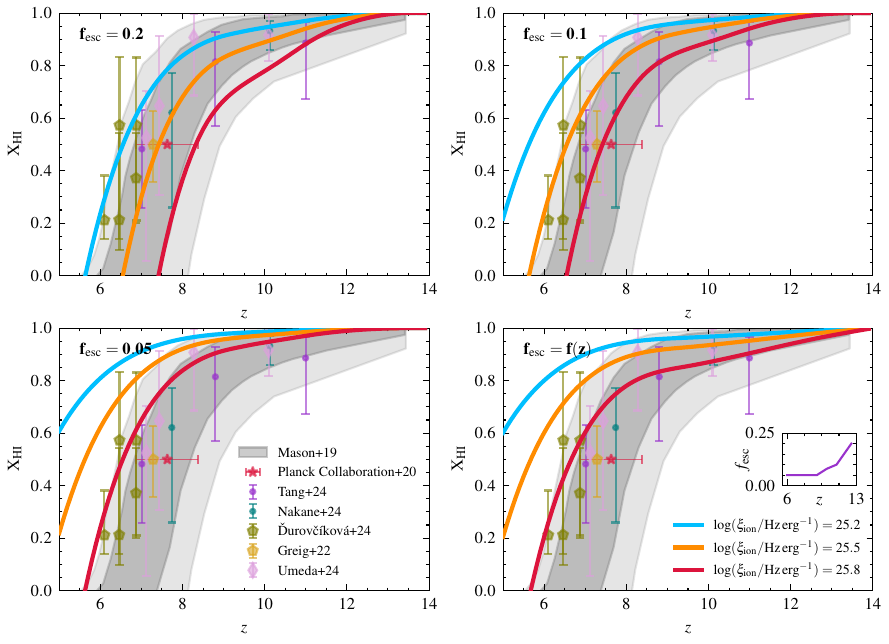}
    \caption{The redshift evolution of the hydrogen neutral fraction of the intergalactic medium with our model predictions given by the solid blue, orange and red lines for $\log(\xi_{\mathrm{ion}}/ \mathrm{Hz\, erg^{-1}})=25.2$, 25.5, 25.8 respectively where the UV LF is integrated to a limit of $M_{\rm UV}=-13$. In each panel the models are computed using different escape fractions with $f_{\rm esc}=0.2$ (upper left), $f_{\rm esc}=0.2$ (upper right), $f_{\rm esc}=0.05$ (bottom left) and for an evolving $f_{\rm esc}$ as function of redshift (bottom right and inset). Observational measurements of the neutral fraction are shown for the CMB \citep{planck2018}, Ly$\alpha$ equivalent widths \citep{tang2024,nakane2024}, QSO damping wings \citep{greig2022,Durovcikova2024} and for LBG damping wings \citep{umeda2024}. The grey shaded region shows the model from \citet{mason2019}(dark grey: 68th percentile; light grey: 95th percentile)}  
    \label{fig:x_HI}
\end{figure*}

Pre-\textit{JWST}, a canonical value of $\xi_{\rm ion}\simeq10^{25.2} \rm Hz\, erg^{-1}$ was typically assumed for reionization calculations \citep[e.g.][]{robertson2015,bouwens2016,matthee2017}. In contrast, initial \textit{JWST} results suggested that high-redshift galaxies displayed higher efficiencies; $\xi_{\rm ion}\simeq10^{25.8} \rm Hz\, erg^{-1}$ \citep[e.g.][]{atek2024,simmonds2024a}. However, more recent {\it JWST}-based studies have tended to revise-down estimates of $\xi_{\rm ion}$, closer
to typical pre-\textit{JWST} values \citep{simmonds2024b,pahl2024,begley2024}. In order to span the range of observational constraints, in our calculations we adopt three different values of $\xi_{\rm ion}$ with $\log(\xi_{\rm ion} / \mathrm{Hz\, erg^{-1}})=25.2, 25.5, 25.8$.

In a similar fashion, we also consider three different values of $f_{\rm esc}$. 
At the highest redshifts where it is possible to directly measure $f_{\rm esc}$ (i.e. $3<z<3.5$), recent studies based on both photometry and spectroscopy have suggested typical values of $f_{\rm esc}\simeq 0.05$ \citep[e.g.][]{pahl2021,begley2022}. At higher redshifts, estimates of $f_{\rm esc}$ are typically based on the relationship between $f_{\rm esc}$ and the UV slope ($\beta$) established at low-redshift \citep{chisholm2022}. \citet{cullen2023} report an average UV slope of $\beta=-2.22$ at $z\simeq10$, corresponding to $f_{\rm esc}=0.07$ using the relationship from \citet{chisholm2022}. They also report an average UV slope of $\beta=-2.6$ at $z\simeq11.8$, indicating an increase in escape fraction with redshift to $f_{\rm esc}=0.19$. Given this range of escape fractions, we determine our model of the neutral fraction as a function of redshift for three fixed values of $f_{\rm esc}=0.05,0.10,0.20$. In addition, we also consider an evolving $f_{\rm esc}$ model, based on the UV slope measurements from \citet{cullen2023}. 

Armed with our calculated values of $\Dot{n}_{\rm ion}$, we determine the neutral fraction of the IGM by solving the \citet{madau1999} differential equation governing the change in the ionised fraction of the universe with time:
\begin{equation}
    \frac{dX_{\rm HII}}{dt} = \frac{\dot{n}_{\rm ion}}{\langle n_{\rm H}\rangle} - \frac{X_{\rm HII}}{t_{\rm rec}}
\end{equation}

\noindent
where $X_{\rm HII}$ is the fraction of ionised hydrogen (and hence the neutral fraction is given by $X_{\rm HI} = 1 - X_{\rm HII}$), $\langle n_{\rm H}\rangle$ is the average comoving number density of hydrogen and $t_{\rm rec}$ is the timescale of recombination. This last parameter is defined by:
\begin{equation}
    t_{\rm rec} = [\mathrm{C_{\rm HII}} \alpha_{\rm B}(T) n_{\rm e}(1+z)^3 ]^{-1}
\end{equation}
Here,  $\alpha_{\rm B}(T)$ is the case-B recombination coefficient, which we assume is $2.59\times10^{-13}\, \rm cm^{3}\,s^{-1}$ given $T=10^4 \, \rm K$, and $C_{\rm HII}$ is the IGM clumping factor, which we assume follows the redshift-dependent form provided by \citet{shull2012}. The resulting evolution of $X_{\rm HI}$ with redshift corresponding to our different parameter choices is shown in Fig.~\ref{fig:x_HI}, along with the latest observational constraints from the literature.

It can be seen from Fig.~\ref{fig:x_HI} that different choices of $f_{\rm esc}$ and $\xi_{\rm ion}$ have a significant impact on the timeline of reionization. 
For $f_{\rm esc}=0.2$, we find that reionization completes by $z=7$, for all but the lowest considered ionizing photon production efficiency ($\log(\xi_{\mathrm{ion}}/ \mathrm{Hz\, erg^{-1}})=25.2$).
In contrast, if we adopt $f_{\rm esc}=0.1$, reionization is only able to complete by $z\simeq 6$ assuming $\log(\xi_{\mathrm{ion}}/ \mathrm{Hz\, erg^{-1}})=25.8$.
If we further reduce the assumed escape factor to $f_{\rm esc}=0.05$, then it can be seen that reionization can only be completed by $z\simeq 6$ for an adopted ionizing photon production efficiency of $\log(\xi_{\mathrm{ion}}/ \mathrm{Hz\, erg^{-1}})\geq25.8$. Finally, using the evolving prescription for $f_{\rm esc}$, reionization also only ends by $z\simeq 6$ for $\log(\xi_{\mathrm{ion}}/ \mathrm{Hz\, erg^{-1}})\geq25.8$ (due to the relatively low value of $f_{\rm esc}$ across most of the relevant redshift range).

\section{Discussion}
\label{sec:discussion}
\subsection{A constant star-formation efficiency}
\label{sec:cst_SF}
We have presented a simple theoretical model which is able to match the observational constraints on both the GSMF at $z=6-8$ and the UV LF at $z=6-13$ (i.e. out to the highest redshifts for which robust measurements of these two key scaling relations currently exist). This was achieved using a redshift-independent form of the stellar-to-halo mass relation and therefore a redshift-independent form of the star-formation efficiency (as a function of halo mass) that is consistent with that observed at $z=0$. We first describe this model in the context of the GSMF and then the UV LF.

\subsubsection{Modelling the GSMF with constant star-formation efficiency}

Firstly,  we tested our constant efficiency model against recent measurements of the evolving GSMF, finding good consistency with the observationally derived data points. The viability of a non-evolving star-formation efficiency is also consistent with several pre-\textit{JWST} measurements of the GSMF obtained through abundance matching \citep{behroozi2013,stefanon2021} as well as the predictions of at least some simulations \citep{feldmann2024}. However, arguments have also been advanced in support of a redshift-dependent star-formation efficiency, based both on  new observations \citep{shuntov2024} and alternative simulations \citep{ceverino2024}. Nonetheless, across most of the mass range, the evolution with redshift that has been suggested has been relatively mild, with even the evolving efficiency functions generally lying  within the scatter of the $z=0$ relation. 

Recently, however, claims have been advanced that the GSMF requires significantly increased star-formation efficiency at $z>6$ for the most massive galaxies. In particular, \citet{shuntov2024} make the dramatic claim that, at the highest redshifts, star-formation efficiency rises to $\epsilon=0.8-1.0$ for $\log(M_{\star}/ \rm M_{\odot})>10.8$. However, this conclusion is based on highly uncertain high-mass data points which are particularly vulnerable to contamination and bias, for example due to incorrect photometric redshifts (especially likely in this study due to the currently limited filter set of the COSMOS-Web imaging). In addition to the positive bias at the high-mass end resulting from even a tiny number of incorrect photometric redshifts, and the inevitability of Eddington bias \citep{eddington1913}, it has also been demonstrated that the use of NIRCam photometry alone can result in the stellar masses of some galaxies being over-estimated by up to $\simeq0.5$ dex \citep{wang2024}. 

Finally, while  the robust measurement of the GSMF is currently limited to $z\leq8$ due to the requirement of sufficiently long wavelength coverage red-ward of the Lyman break, we note that this has not deterred several authors from providing measurements at $z = 9-11$, which have again in some cases been used to argue for increased star-formation efficiency \citep{harvey2024,weibel2024,shuntov2024}. 
Although these tentative estimates are not shown in Fig.~\ref{fig:SMF}, we have checked that, within the observational uncertainties, our model does in fact successfully reproduce the proposed form of the GSMF  at $z=9$ and $z=11$.

We therefore conclude that, at present, the available observational constraints on the evolving 
GSMF provide no clear evidence for a significant increase in star-formation efficiency in massive galaxies at $z\ge 6$.

\subsubsection{The cosmic stellar mass density at $z=6-13$}
\begin{figure}
	\includegraphics[width=\columnwidth]{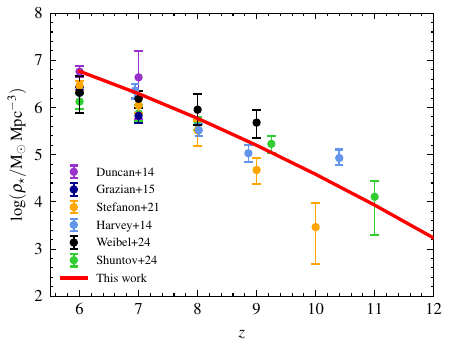}
    \caption{The prediction of our constant star-formation efficiency model for the evolving cosmic stellar-mass density (CSMD), $\rho_{\star}$, from $z=6-12$ (red). The model prediction is in good agreement with observational constraints from \citet{duncan2014,grazian2015,stefanon2021,harvey2024,weibel2024,shuntov2024}. Where necessary, all observational data points have been converted to a \citet{chabrier2003} initial mass function and are integrated over the mass range $10^{8} - 10^{13}\,\rm M_{\odot}$.}  
    \label{fig:CSMD}
\end{figure}

A useful global property that can be determined from the GSMF is the evolving cosmic stellar-mass density, $\rho_{\star}$, which describes the stellar mass per unit co-moving volume as a function of redshift. This is determined by the mass-weighted integral of the GSMF given by:
\begin{equation}
    \rho_{\star} = \int^{M_2}_{M_1} M_{\star}\phi(M_{\star},z)dM_{\star}
\end{equation}

\noindent
where stellar mass is given by $M_{\star}$, the GSMF is given by $\phi(M_{\star},z)$ and $M_1$, $M_2$ define the chosen integration limits. Here we adopt $M_1=10^8\,\rm M_{\odot}$ and $M_2=10^{13}\,\rm M_{\odot}$ as typically assumed in observational studies of the GSMF. In Fig.~\ref{fig:CSMD} we show the resulting prediction of our model (red line) in comparison to the observational measurements from \citet{duncan2014,grazian2015,stefanon2021,harvey2024,weibel2024,shuntov2024}. This demonstrates that our constant star-formation efficiency model is able to successfully reproduce the observed evolution of $\rho_{\star}$ out to $z\geq11$ while also highlighting the significant scatter of the current observational determinations. Fig.~\ref{fig:CSMD} thus reaffirms that an increase in star-formation efficiency at early times is {\it not} required to explain the growth of cosmic stellar-mass density at early times.

\subsubsection{Star-formation efficiency and the UV LF at $z=6-13$?}
Using a similar type of model to the one considered here, \citet{harikane2023a} concluded that the assumption of constant star-formation efficiency was able to reproduce the UV LF out to $z\simeq10$, but that an increased efficiency was required at still higher redshifts. There have also been a number of other studies suggesting that increased star-formation efficiency is required to explain the form of the UV LF at $z > 10$, especially at the bright end \citep[e.g.][]{dekel2023}. 

Of course, the difficulty in using the UV LF as a measurement of the star-formation efficiency is that the UV luminosity of a galaxy does not simply depend on its stellar mass. Even at early times, a galaxy of a given stellar mass can have a range of UV luminosities dependent on its star-formation history and the age/metallicity of the stellar population that dominates the rest-frame UV. Indeed, in the simple model discussed here we have demonstrated that the assumption of constant star-formation efficiency can be maintained by appealing to ever younger stellar populations to produce an increasing average UV luminosity with increasing redshift beyond $z \simeq 8$. It is only through the flexibility of decreasing typical age at extreme redshifts that our constant-efficiency model is able to explain the UV LF at $z=10-13$. 

Somewhat surprisingly, the idea of decreasing characteristic age at extreme redshifts is not regularly cited as an explanation for the observed modest evolution of the UV LF (and hence $\rho_{\rm SFR}$) from $z \simeq 8$ out to $z \simeq 13$. Nevertheless, at some stage young ages become inevitable as we are rapidly running out of time at these redshifts, and indeed our approach is vindicated by the observed consistency of our model (without modification), with measurements of the GSMF (which offers a more direct probe of star-formation efficiency). A key prediction resulting from this trend of younger ages with increasing redshift, is that this trend can continue only for so long as at a certain point, galaxies can no longer be made younger. In fact, the ages at all redshifts in this study are consistent with a formation redshift of $z\simeq15$, indicating that at $z\gtrsim15$ there should be a more rapid decline in $\rho_{\rm SFR}$ due to the rapid evolution of the halo mass function.

Finally, we note that while our simple constant-efficiency model can reproduce the UV LF at $z>10$, more extensive spectroscopy of galaxies at $z >10$ is required to investigate the physical properties of early galaxies in detail, including the ages of their stellar populations. To date, only a handful of galaxies at such extreme redshifts (extending up to $z = 14.2$) have been observed with NIRSpec \citep[e.g.][]{bunker2023,arrabalharo2023,castellano2024} and ALMA \citep{schouws2024,zavala2024}, and so it remains unclear how typical their properties are of the wider emerging galaxy population at these early times. Fortunately, in Cycle 3 \textit{JWST} is set to observe an order-of-magnitude more galaxies at $z>10$ with NIRSpec, through a series of major spectroscopic surveys (e.g. CAPERS GO 6368). 

\subsection{No evidence for removing feedback in high-mass halos}
A number of studies have presented evidence for an excess of bright galaxies ($M_{\rm UV}\lesssim-21$) at $z\geq6$ compared to expectations based on the assumption (well justified at lower redshifts) that the LF follows the exponentially declining form of a Schechter function \citep[e.g.][]{bowler2014,bowler2020,stefanon2019}. 

One possible explanation for this phenomenon is the weakening or removal of high-mass feedback at early times. With feedback reduced or even non-existent at high masses, galaxies can rapidly assemble larger stellar masses and therefore produce greater UV luminosities, thus potentially explaining the observed slow evolution in the bright-end of the UV LF. High-mass feedback is typically explained by the impact of AGN in massive galaxies whereby their jets expel gas from the galaxy therefore quenching further star formation, potentially permanently  \citep{hopkins2006}. Alternatively, halo quenching can also act as a feedback mechanism, with the shock-heating of gas in massive halos effectively removing the reservoir of cold gas required for star formation \citep[][]{somerville2008,gabor2015}.

It might reasonably be expected that both of these high-mass feedback mechanisms would be less effective at extreme redshifts, with super-massive black holes and high mass dark-matter halos requiring time to grow.
However, as we discussed in, Section~\ref{sec:submm} there exists no clear evidence for the removal of high-mass feedback at $z\geq6$ given the current observational constraints on the evolving GSMF.

This motivated our test of the sub-mm galaxy counts using our model of the intrinsic form of the UV LF. This is a useful test of the high-mass end of star-formation efficiency because it is now well established that the most massive star-forming galaxies are heavily obscured by dust and are thus unlikely to feature even at the bright end of the UV LF (despite obviously still appearing at the high-mass end of the GSMF). We have found that the number of bright high-redshift galaxies observed in current sub-mm surveys \citep[e.g.][]{simpson2019,mckinney2024} also provides little evidence for any decline in high-mass feedback at early times, with the observed prevalence of bright sub-mm galaxies ($\simeq 1$ per square degree at $z > 5.5$ forming stars at a rate $\ge 300 {\rm M_{\odot} yr^{-1}}$) being completely consistent with the predictions of our fiducial model. Furthermore, in Section~\ref{sec:dust_attenuation} we demonstrated that we obtain physically plausible relationships between the UV attenuation and stellar mass for $z\geq6$ (discussed further in Section~\ref{sec:dust_discussion}) with an unchanged high-mass star-formation efficiency. 

Although in our model the bright end of the UV LF can be fully explained by the level of dust attenuation, and the sub-mm galaxy counts provide some further constraining power, the precise differentiation between weakening star-formation efficiency and increasing dust attenuation will require further observational measurements. Direct measurements of the dust content of galaxies at $z\geq6$ with ALMA and NOEMA will help determine the relationship between stellar mass, UV luminosity and dust attenuation in these galaxies. This requires both sufficient depth (to be able to detect or robustly constrain the dust continuum) and surveys covering sufficient area in order to detect a statistically significant number of galaxies with stellar masses of $\log(M_{\star}/ \mathrm{M_{\odot}})>10$ at $z\geq6$.

While our tests of the high-redshift sub-mm galaxy counts do not provide any strong support for the removal of high-mass feedback at early times, currently we cannot precisely determine the normalisation and form of the high-mass end of the stellar-to-halo mass relation at $z\geq6$. This would require improved measurements of the high-mass end of the GSMF which can only be enabled by yet wider-area NIRCam surveys with \textit{JWST}. 

Progress is being made in this direction. To date \citet{harvey2024} have presented measurements of the GSMF using $\simeq187$ sq. arcmin of NIRCam imaging, while  \citet{weibel2024} used a larger area of $\simeq500$ sq. arcmin from PRIMER \citep{dunlop2021}, JADES \citep{eisenstein2023} and CEERS \citep{bagley2023}. More recently, \citet{shuntov2024}
have presented measurements of the GSMF based on the Cosmos-Web program \citep{caset2023b} which significantly expands the available survey area to $\simeq 0.5$ sq. degrees.
However, as discussed above, the COSMOS-Web survey is currently limited to only 4 NIRCam filters (F115W, F150W, F277W, F444W) which results in significantly less robust photometric redshifts and estimated stellar masses. The upcoming {\it JWST} Cycle-3 COSMOS-3D program (PI: Kakiichi) will provide F200W and F356W imaging across $\sim0.33$ sq. degrees of the COSMOS-Web footprint therefore providing contiguous NIRCam wavelength coverage from $1-4.4 \mu$m which should provide significantly improved stellar mass constraints for the galaxies detected in this field. 

Finally, another important consideration is the wavelength coverage of the observations.  Interestingly, \citet{wang2024} used MIRI imaging at $7.7\mu$m to extend the wavelength coverage used to measure galaxy stellar masses beyond that which can be accessed with NIRCam. They concluded that this has an important impact in the sense that the inclusion of the MIRI data results in a significant reduction in the observed number densities of massive galaxies. Therefore, the robust determination of the GSMF at $z>6$ will require further expansion of the existing  {\it JWST} imaging surveys, ideally with substantial NIRCam and MIRI overlap (as was designed for PRIMER).

\subsection{The evolution of the UV attenuation - stellar mass relation}
\label{sec:dust_discussion}
Our model-derived form of the $A_{1500}-M_{\star}$ relations at $z\geq6$ as shown in Fig.~\ref{fig:dust_mass} appear to show some evidence for some modest redshift evolution, with a slightly lower normalisation than seen in the $z=2$ relation derived by \citet{mclure2018}. However, the evidence for a change is not compelling, and perhaps more interesting is the extent to which the derived relations still appear to mirror the form of the $z \simeq 2$ relation out to the highest redshifts. This might be regarded as surprising, given that a (relative) lack of dust attenuation in bright ($M_{\rm UV}\lesssim-21$) galaxies at $z>10$ has been suggested as an explanation for their observed luminosity \citep[e.g.][]{ferrara2023}. Moreover, there is also growing evidence from the analysis of the UV continuum slopes exhibited by extreme-redshift galaxies for a transition towards very little or even zero dust attenuation at $z \geq  11$ \citep[e.g.][]{topping2023,morales2023,cullen2023}. 

However, it transpires that the galaxies uncovered to date at $z \geq 11$ cannot yet settle the issue of whether dust is present or absent at such early times. This is because, as can be seen in Fig.~\ref{fig:UVLF}, the brightest bin sampled in our analysis of the UV LF at $z \simeq 11$ lies at $M_{\rm UV}=-21.25$ which, within our model, corresponds to a stellar mass of $10^{8.4}\, \rm M_{\odot}$. As can be seen from Fig.~\ref{fig:dust_mass}, at such modest stellar masses (and at all lower stellar masses) negligible dust extinction would be expected even at $z \simeq 2$. As a result, the observed UV LF shown in Fig.~\ref{fig:UVLF} is actually consistent with both dusty and dust-free models. In reality, the current observational constraints on dust attenuation in individual galaxies at $z\geq11$ remain sparse. For example, a number of spectroscopically confirmed galaxies at $z\geq11$ indicate negligible levels of dust attenuation at $M_{\rm UV}\leq-20$ \citep[e.g.][]{bunker2023,castellano2024}, but spectroscopic observations of a galaxy at $z=14.3$ hint at some modest (albeit highly uncertain) obscuration \citep[$A_{\rm V}\simeq0.3$;][]{carniani2024}. Wider area surveys, still deep enough to detect galaxies at $z \simeq 11$, are thus required to properly sample the bright end of the UV LF at these redshifts, uncovering rarer, more massive galaxies. Only then will it be possible to determine 
whether dust remains prevalent in higher-mass galaxies at $z \simeq 11$ and indeed whether the $A_{1500}-M_{\star}$ relation at these early times is significantly different from that seen at cosmic noon.

A number of other theoretical models have been developed which attempt to predict the dust properties of galaxies at $z>10$. For example, \citet{ferrara2023,ferrara2024} suggest that although moderate levels of dust are produced, this dust is removed by radiation-driven outflows. They suggest that such dust removal is required to match the blue colours of early galaxies due to the compact nature of galaxies at $z\geq11$, as a result of which even a relatively small mass of dust might be expected to produce significant attenuation. Recent ALMA and NOEMA observations of GS-z14 and GN-z11 respectively have attempted to directly determine the dust mass through measurement of the dust continuum in the sub-mm \citep{schouws2024,carniani2024b, fudamoto2024}. However, these measurements lacked sufficient depth to exclude dust at the level implied by the current models. Significantly deeper ALMA observations are therefore required to directly constrain the levels of dust in galaxies at $z\geq11$.

\subsection{Is there an ionizing photon budget crisis?}
Recent \textit{JWST} results have been interpreted by some as suggesting that there may be an ionizing photon budget `crisis' \citep{munoz2024}, in the sense that too many ionizing photons are emitted from the galaxy population at $z>6$ to be consistent with other constraints on the progress of cosmic hydrogen reionization. 
This suggestion has arisen in part due to assumptions made regarding the values of the three parameters contained in Equation \ref{eq:nion}, to which the inferred reionization history is sensitive. 

\textit{JWST} has unlocked the ability to measure $\xi_{\rm ion}$ at $z>6$ for the first time. Early \textit{JWST} results indicated that $\xi_{\rm ion}$ increased as a function of redshift, perhaps reaching significantly higher values of $\log(\xi_{\mathrm{ion}}/ \mathrm{Hz\, erg^{-1}})\simeq26$ at $z>6$ \citep{simmonds2024a,atek2024}. This was in part justified by an expectation that the increased burstiness of star formation at early times would result in an increase in the production of ionizing photons. However, recent deeper surveys have led to lower inferred values more typical of what had been established at lower redshifts pre-\textit{JWST}, with various studies now concluding in favour of a more mild evolution with redshift and an average efficiency of $\log(\xi_{\mathrm{ion}}/ \mathrm{Hz\, erg^{-1}})\simeq25.3$ \citep{simmonds2024b,pahl2024,begley2024}. This change in stance may be explained by the fact that the galaxies in earlier studies were selected due to the strengths of key emission lines (e.g. H$\alpha$, [OIII]) and therefore only provided measurements of $\xi_{\rm ion}$ for the subset of galaxies experiencing the strongest bursts of star formation.

The second factor that led to the suggestion of an ionizing photon budget crisis is the assumed value of the typical escape fraction for ionizing photons, $f_{\rm esc}$. For example,  \citet{munoz2024} assume an initial value of $f_{\rm esc}=0.2$ simply because this was the value adopted in \citet{robertson2015} as that which was required to complete reionization by $z\simeq5.5$ (given the other constraints available at that time). Using $f_{\rm esc}=0.2$, as shown in the upper left panel of Fig.~\ref{fig:x_HI}, we find consistency with the suggestion from \citet{munoz2024} that this {\it can} result in an excess of ionizing photons leading to premature reionization, with the highest $\xi_{\rm ion}$ model completely reionizing the hydrogen in the IGM by $z\simeq7.5$. However, pre-\textit{JWST} observational measurements of $f_{\rm esc}$ indicated a significantly lower typical value of $f_{\rm esc}\simeq0.05$, based on attempted measurements at lower redshifts  \citep{pahl2021,saldanalopez2022,begley2022}. Using the $\beta - f_{\rm esc}$ relation from \citet{chisholm2022} in combination with recent measurements of $\beta$ up to $z\simeq13$ from \citet{cullen2023}, we therefore also calculate the predictions of the model assuming an evolving $f_{\rm esc}$ in which (as implied by the new UV slope measurements) $f_{\rm esc}\simeq0.2$ at $z=12$, but then rapidly decreases to $f_{\rm esc}\simeq0.07$ by $z=10$ (shown in the inset panel in the bottom right of Fig.~\ref{fig:x_HI}). However, this change in fact results in a tension with observational constraints on the neutral fraction for even modestly large values of the ionizing production efficiency ($\log(\xi_{\mathrm{ion}}/ \mathrm{Hz\, erg^{-1}})=25.5$). This is because, based on the values inferred from the evolution of UV continuum slopes, the escape fraction only rises to $f_{\rm esc}=0.2$ at $z>10$ and therefore the majority of the reionization history occurs while galaxies have significantly lower escape fractions ($f_{\rm esc}\simeq0.05$). However, we stress that this tension is in the opposite direction to that proposed by \citet{munoz2024}, as this model actually lacks the number density of ionising photons required to complete reionization by $z\sim6$.

\begin{figure}
	\includegraphics[width=\columnwidth]{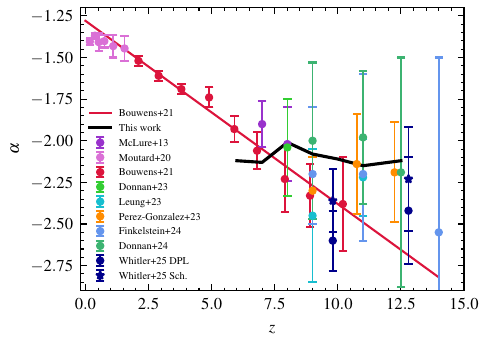}
    \caption{The redshift evolution of the faint-end slope, $\alpha$, of the galaxy UV LF. The solid red line shows the model from \citet{bouwens2021} whereas the solid black line shows the faint-end slope predicted by our model as measured over the luminosity range corresponding to $-18.5<M_{\rm UV}<-17.5$. We also plot a range of observational measurements from  \citet{mclure2013,moutard2020,bouwens2021,donnan2023a,leung2023,perezgonzalez2023,finkelstein2024,donnan2024} and \citet{whitler2025}. Clearly direct observational constraints on $\alpha$ currently remain very weak, but our model predicts that $\alpha$ is expected to plateau around a value of $\alpha \simeq 2$ at extreme redshifts, and should never approach the extremely steep values inferred from extrapolation of pre-{\it JWST} data (as indicated by the red line).}  
    \label{fig:alpha}
\end{figure}

The final parameter that can lead to a predicted excess in the number of ionizing photons is the UV luminosity density, $\rho_{\rm UV}$, produced by the emerging population of young galaxies. This concern was raised in part due to the excess in the observed UV LF compared to some pre-\textit{JWST} models at $z>10$ \citep[e.g.][]{mason2015,yung2019}. However, our model, which closely reproduces the observed UV LF and $\rho_{\rm UV}$ at $z=6-13$ shows that, in fact, this is not in tension with observational measurements of the neutral fraction including the (now well constrained) end-point of reionization \citep[at $z\simeq5.5$;][]{bosman2021}. The calculation of $\rho_{\rm UV}$ is highly sensitive to the limit of the luminosity-weighted integral and hence also to the steepness of the faint-end slope of the UV LF, $\alpha$. Prior to the launch of \textit{JWST}, a range of data provided evidence for a monotonic steepening in $\alpha$ with increasing redshift. In \citet{bouwens2021}, this evolution was described by a linear relationship with redshift, specifically $\alpha = -1.94-0.11(z-6)$ (see Fig.~\ref{fig:alpha}). Extrapolation of this linear relationship then provided a prediction of very steep faint-end slopes at $z>10$. Indeed, assuming this linear extrapolation means that by $z=11$ one would anticipate a faint-end slope of $\alpha\simeq-2.5$ \citep{bouwens2023a}, significantly steeper than is observed at $z\simeq8$ ($\alpha\simeq-2.1$). 

Early \textit{JWST} measurements of the UV LF were unable to constrain $\alpha$ given the initial lack of ultra-deep NIRCam imaging. However, more recent measurements using the JADES \citep{eisenstein2023} and NGDEEP \citep{bagley2024} surveys have enabled a first meaningful attempt at this measurement, yielding $\alpha \simeq -2.2$ to $\alpha \simeq -2.4$ at $z=11$, albeit still with significant uncertainties \citep{leung2023,perezgonzalez2023,donnan2024,whitler2025}. This suggests a significant softening in the evolution in $\alpha$ with increasing redshift compared to some of the aforementioned pre-\textit{JWST} predictions and indeed a plateauing of the faint-end slope around a value of $\alpha \simeq -2.1$ is a clear prediction of our model. This is demonstrated in Fig.~\ref{fig:alpha} where we compare the results of simple power-law fits the the faint end of our evolving model UV LF with existing observational constraints on $\alpha$ at extreme redshifts. 

Our model also predicts that the faint-end slope should actually reduce in steepness as we move to very faint magnitudes as a result of the non-linear faint-end form of the UV LF (resulting from the underlying form of the halo mass function and the assumed ever decreasing star-formation efficiency at lower masses). This is quantified in Table.~\ref{tab:faint_end}, where we tabulate the faint-end slope resulting from simple power-law fits to our evolving model UV LF over the luminosity ranges corresponding to  $-18<M_{\mathrm{UV}}<-16$ and $-17<M_{\mathrm{UV}}<-15$. There is some observational evidence for this shape at $z\simeq2-4$ \citep{parsa2016} but deeper \textit{JWST} imaging is required to measure this at the highest redshifts.
\begin{table}
	\centering
	\caption{The faint-end slope, $\alpha$, of our UV LF model at $z=6-12.5$, derived by fitting a a power-law functional form over a chosen luminosity range. The first column states the redshift of the UV LF. The second column gives the best-fitting faint-end slope to the model obtained by fitting a power-law over the UV magnitude range $-18<M_{\mathrm{UV}}<-16$. The final column then gives the best-fitting faint-end slope derived by fitting to the slightly fainter luminosity range corresponding to  $-17<M_{\mathrm{UV}}<-15$. It is notable that, in our model, the faint-end slope plateaus around $\alpha \simeq -2$ at the highest redshifts, and actually flattens slightly as one moves towards very faint luminosities.}
	\label{tab:faint_end}
    \setlength{\tabcolsep}{4pt} 
	\renewcommand{\arraystretch}{1.15} 
	\begin{tabular}{lcc} 
		\hline
          z & $\alpha$ & $\alpha$\\
		 & $-18<M_{\mathrm{UV}}<-16$  & $-17<M_{\mathrm{UV}}<-15$ \\
		\hline
  6 & $-1.72$ & $-1.66$\\
  7 & $-1.78$ & $-1.71$\\
  8 & $-1.84$ & $-1.78$\\
  9 & $-1.93$ & $-1.87$\\
  10 & $-2.00$ & $-1.93$\\
  11 & $-2.04$ & $-1.97$\\
  12.5 & $-2.16$ & $-2.08$\\
  \hline
	\end{tabular}
\end{table}

Our predicted milder evolution in $\alpha$ with increasing redshift {\it and} the implied decreasing steepness ({\it i.e.} curvature) of the UV LF as one moves to fainter UV magnitudes both limit the number density of faint galaxies able to contribute to reionization. This, coupled with the latest \textit{JWST} constraints on the escape fraction and ionizing photon production means that, in this study, we find no evidence for an excess of ionizing photons. Rather, as discussed above, we find that many of our models lack the number density of ionising photons needed to reproduce  current  observational constraints on the progress of cosmic hydrogen reionization unless we adopt a higher escape fraction ($f_{\rm esc}=0.2$) than required in other studies which assumed steeper faint-end slopes for the LF \citep[e.g.][]{whitler2025}. 

\begin{figure}
	\includegraphics[width=\columnwidth]{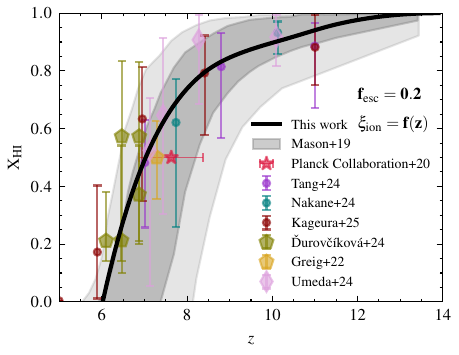}
    \caption{The redshift evolution of the neutral fraction of the intergalactic medium with our final model prediction given by the solid black line where $\log(\xi_{\mathrm{ion}}/ \mathrm{Hz\, erg^{-1}})=0.035 z + 25.08$ \citep{begley2024} and the UV LF is integrated to a limit of $M_{\rm UV}=-13$. We assume a fixed $f_{\rm esc}=0.2$. Observational measurements of the neutral fraction are shown for the CMB \citep{planck2018}, Ly$\alpha$ equivalent widths \citep{tang2024,nakane2024,kageura2025}, QSO damping wings \citep{greig2022,Durovcikova2024} and for LBG damping wings \citep{umeda2024}. The grey shaded region shows the model from \citet{mason2019} (dark grey: 68th percentile; light grey: 95th percentile)}  
    \label{fig:final_X_HI}
\end{figure}

In Fig.~\ref{fig:final_X_HI} we show the final prediction of our model for the redshift evolution of the neutral hydrogen fraction. Here we use the evolving value of $\rho_{\rm UV}$ calculated from the luminosity-weighted integration of our new model UV LF down to $M_{\rm UV}=-13$, assume an escape fraction of $f_{\rm esc}=0.2$ across the full relevant redshift range, and adopt the redshift-dependent form for $\xi_{\rm ion}$ deduced by \citet[][]{begley2024}. The resulting predicted progress of reionization is clearly consistent with the other available observational constraints plotted in Fig.~\ref{fig:final_X_HI}. Crucially, while reionization is predicted to start at quite early times, the model delivers on the now well-established requirement that hydrogen reionization should terminate at $z\simeq5.5$.

\section{Conclusions}
\label{sec:conclusions}

We have developed and tested a simple theoretical model of galaxy evolution with a constant star-formation efficiency, designed in part to check the veracity of recent claims that enhanced star-formation efficiency is required at early cosmological times in order to explain the prevalence of UV luminous galaxies at extreme redshifts, as revealed by {\it JWST}.

We show that this model, with no free parameters other than mass-light ratio, can indeed perfectly reproduce both the shape and normalisation of the galaxy UV luminosity function (LF) at $z \simeq 11$, and indeed can match the evolving UV LF over the entire redshift range $z \simeq 6 -13$, provided we: i) allow the typical age of the galaxy stellar populations to decrease monotonically from $\simeq 130$\,Myr at $z = 6$ to $\simeq 10$\,Myr at $z \simeq 13$, and ii) invoke dust obscuration to explain the form of the UV LF at luminosities brighter than $M_{UV} \simeq -20$ for redshifts $z \leq 10$.

We have also compared the predictions of our model against the latest observations of the evolving galaxy stellar mass function (GSMF) over the redshift range $z \simeq 6 - 8$. The requirement to fit both the UV LF and the GSMF breaks the degeneracy between mass-to-light ratio and star-formation efficiency and we find that our model is consistent with current observational constraints on the GSMF at these epochs, {\it without alteration or tuning}. This provides reassurance that our assumed non-evolving efficiency is completely consistent with recent {\it JWST} observations of the young Universe.

We have also explored a modification of our model involving the removal of high-mass feedback at early times which would increase the number of highly star-forming galaxies present in sub-mm surveys. However, the predictions of our {\it unmodified} model are more consistent with existing constraints from sub-mm surveys, which reveal typically only $\simeq 1$ sub-mm galaxy per square degree with a star-formation rate $\ge 300\,{\rm M_{\odot} yr^{-1}}$ at $z \ge 6$. 

Finally, we have explored the implications of our model for the predicted progress of cosmic hydrogen reionization, in part to test the validity of some recent claims that the latest \textit{JWST} observations suggest early galaxies may emit too many ionizing photons to be consistent with other constraints on reionization. We find that, using the latest measurements of $f_{\rm esc}$ and $\xi_{\rm ion}$ in combination with our model UV LFs, our predictions are not in tension with observational measurements of the timeline of reionization. 
Our model also predicts that the faint-end slope of the UV LF should plateau at $\alpha \simeq -2.1$ beyond $z \simeq 8$, rather than continuing to steepen at higher redshifts (as often predicted before the advent of {\it JWST}), a prediction which helps to mitigate any potential ionizing photon crisis.

Because we invoke ever-declining stellar ages (rather than increased star-formation efficiency) to offset the decline in the underlying halo mass function when modeling the extreme-redshift evolution of the UV LF, another clear and {\it prediction} of our model is that there should be a more rapid drop-off in the galaxy number density galaxies at $z \simeq 15$. It is therefore interesting to note that \citet{donnan2024} presented tentative evidence for a relatively rapid drop-off in $\rho_{UV}$ at $z \simeq 14.5$ and indeed a compelling galaxy candidate at $z \ge 15$ has yet to be discovered, despite intensive efforts with ever deeper {\it JWST} data. Time will tell whether $z \simeq 15$ really is the long-sought epoch of galaxy formation, as predicted by our model.

\section*{Acknowledgements}
We thank the referee for their valuable comments which have improved the quality of the manuscript.
C.\,T. Donnan, D.\,J. McLeod, R.\,J. McLure, and J.\,S. Dunlop acknowledge the support of the Science and Technology Facilities Council (STFC). J.\,S. Dunlop also acknowledges the support of the Royal Society through a Royal Society Research Professorship.
F. Cullen acknowledges support from a UKRI Frontier Research Guarantee Grant [grant reference EP/X021025/1].
\section*{Data Availability}

The data is available from the authors upon reasonable request.



\bibliographystyle{mnras}
\bibliography{LF_model} 







\bsp	
\label{lastpage}
\end{document}